\documentclass[11pt]{article}

\usepackage[utf8]{inputenc}
\usepackage[T1]{fontenc}
\usepackage{lmodern}

\usepackage{geometry}
\usepackage{setspace}
\usepackage{graphicx}
\usepackage{amsmath,amssymb}
\usepackage{booktabs}
\usepackage{url}
\usepackage{hyperref}
\usepackage[authoryear]{natbib}

\hypersetup{
  colorlinks=true,
  linkcolor=black,
  citecolor=black,
  urlcolor=black
}

\title{\textbf{Clarifying Core Dimensions in Digital Maturity Models:\\
An Integrative Approach}}

\author{
Eduardo C. Peixoto\thanks{Corresponding author. Email: \texttt{ecp@cesar.school}} \\
\small CESAR School, Recife, PE, Brazil
\and
Hector Oliveira \\
\small CESAR School, Recife, PE, Brazil
\and
Geber L. Ramalho \\
\small Federal University of Pernambuco (UFPE), Recife, PE, Brazil
\and
Cesar França \\
\small Federal Rural University of Pernambuco (UFRPE), Recife, PE, Brazil
}

\date{} 

\begin{document}
\maketitle

\begin{abstract}
Digital Transformation (DT) initiatives frequently face high failure rates, and while Digital Maturity Models (DMMs) offer potential solutions, they have notable shortcomings. Specifically, there is significant disparity in the dimensions considered relevant, a lack of clarity in their definitions, and uncertainty regarding their components. This study aims to provide a clearer understanding of DMMs by proposing integrative definitions of the most frequently used dimensions. Using a Systematic Mapping approach, including automatic search and snowballing techniques, we analyzed 76 DMMs to answer two Research Questions: (RQ1) What are the most frequent dimensions in DMMs? and (RQ2) How are these dimensions described, including their components? We reconcile varying interpretations of the ten most frequent dimensions — Organization, Strategy, Technology, Culture, Process, Operations, People, Management, Customer, and Data — and propose integrative definitions for each. Compared to previous analyses, this study provides a broader and more recent perspective on Digital Maturity Models.
\end{abstract}

\vspace{0.5em}

\section{Introduction}
\label{sec1}

The maturity of digital technologies have enabled business infrastructures that turned operating and transaction costs lower, allowing the rise of new and more competitive business models. Nowadays, to remain on the market and defend against new entrants, being digital has become an imperative to all companies \citep{boulton2018disruption}.
Nevertheless, transforming companies’ strategy and operations from analogical to the digital world is not straightforward. They face (to say a few) challenges on culture, talents, customers and competition. According to some articles, investments in Digital Transformation (DT) reached US\$ 2.49 trillion in 2024 (only in technology and services), but 70\% of all DT initiatives fail \citep{tabrizi2019digital}.

To assess and evaluate a company’s progress towards a high-level state of a process, practitioners frequently follow referential maturity models. In this sense, a concept closely linked to DT is the concept of Digital  Maturity (DM). According to \citet{Aslanova2020}, DM is the final stage of DT, which companies aspire to achieve. The maturity models that specifically gauge the state of DT are called Digital Maturity Models (DMM).

Although DMMs are widely used by practitioners, they are far from perfect and therefore require more attention from scholars. The first issue is that  there is no consensus among the current DMMs on which dimensions are important to the DT process. \citet{teichert2019digital} found a significant disparity across models, with 125 distinct dimensions identified in 22 models, but only 41 being common across all. Second, many DMMs suffer from poor definition of dimensions, lacking consistency and clarity in what is being measured, as also noted by \citet{teichert2019digital}.

Finally, the components of each dimension are also not well identified. These issues are related: the lack of clarity probably leads to variations in how dimensions are understood and applied, contributing to the disparity among models and complicating any attempt at comparative analysis. 

In the current literature, there are five systematic analyses (reviews or mappings) that scrutinize DMMs publications on multiple economic sectors, as discussed in Section \ref{sec6}. However, they do not address the problem of well defining the dimensions or their components. Furthermore, they cover a limited number of DMMs and/or a limited application scope. 

In this study, we aim to provide an integrative and coherent view of DMMs by systematically discussing and explicitly defining their most frequently used dimensions. To achieve this objective, we conducted a systematic mapping of the literature. Following the screening and selection of studies retrieved through automated searches and snowballing techniques, a total of 76 DMMs were identified and analyzed to address the research questions (RQ) presented below.

\begin{itemize}
    \item RQ1 - What are the most frequent dimensions in DMM?
    \item RQ2 - How are these dimensions described,  including which are their components?
\end{itemize}

The list of DMMs, in \ref{app2}, is more extensive and up-to-date (up to 2023) than previous studies.
By clarifying and reconciling a central element in DMMs, their dimensions, we promote a better understanding of DMMs, thereby enabling more effective management of the DT process by practitioners and fostering the development of more robust and effective models by scholars.

\section{Theoretical Background}
\label{sec2}

\subsection{DT, What is It?}  \label{subsec1}
In the early 2010s, \citet{andreessen2011software} highlighted the growing role of software and digital platforms in transforming business models across various sectors, a trend that would later be recognized as DT. \citet{rogers2016digital} emphasized that the mass adoption of the internet enabled new business approaches by reshaping customer preferences, support infrastructure, and business ecosystems, leading to the rise of innovative companies and operational models. According to \citet{boulton2018disruption}, once digital transactions account for more than 20\% of total market transactions, it becomes impossible to compete without adopting digital strategies.

While the word digital in the term DT may have broad consensus, meaning digital technologies (e.g., social, mobile, artificial intelligence, cloud, and internet of things), the word transformation has much heavier implications and cannot be confused with any type of change, transition, technological innovation, innovation, or process improvement, as discussed by \citet{morakanyane2017conceptualizing}. Transformation means a change in form, appearance or structure to create something new that has never existed before and needs to meet the criteria to be big, bold and lead to better results \citep{gong2021developing}.

There are several literature analyses trying to unify the concept of DT. \citet{morakanyane2017conceptualizing} sought to reconcile several definitions to propose the DT as “An evolutionary process that leverages digital capabilities and technologies to enable business models, operational processes and customer experiences to create value”. \citet{gong2021developing} presented a broader definition: “A process of fundamental change, enabled by the innovative use of digital technologies accompanied by the strategic leveraging of key resources and capabilities, with the aim of radically improving an entity and redefining its value proposition for its stakeholders”.

\subsection{Digital Maturity, What is It?} \label{subsec2}
The notion of maturity is used to define, assess and form a guideline and a basis for evaluating the progress in business, i.e., the maturity of process or a technology \citep{gokalp2017development}. In DT, organizations often turn to existing digital maturity frameworks \citep{valdez2016digital}, as they provide a structured approach to understanding and addressing the challenges. In this sense, a concept closely linked to DT is the concept of Digital Maturity. We can say that Digital Maturity is the final or perfect stage of DT, which companies aspire to achieve \citep{Aslanova2020}.

By assessing their level of Digital Maturity, organizations can measure how well they have integrated digital technology into their operations and their readiness to continuously adapt consistently to ongoing digital change. Then, measuring Digital Maturity is not just about tracking technology adoption or project completion. It is about understanding the impact of these efforts on business, making informed strategic decisions, promoting a culture of continuous improvement and, ultimately, ensuring the adaptation and competitiveness of companies that have established themselves and thrived in physical infrastructures towards a new networked digital business context \citep{thordsen2020measure, siedler2021maturity, canetta2018development}. 

\subsection{DMMs, What are They?} \label{subsec3}
According to \citet{schumacher2016maturity}, Maturity Models are extensively applied to assess companies' progression towards advanced production processes. The underlying supposition of using Maturity Models is that, as the degree of maturity becomes higher, better progress is achieved in different aspects that contribute to the maturation of the process \citep{agca2017industry}.

Throughout a DT journey, managers need to know the current state of their organization with regard to the transformation process. They need to understand how companies actually face the transformation, how to approach their organization transformation\citep{hess2016options}, and ultimately what would make the transformation successful \citep{heckmann2016organizational}. DMMs provide some guidance in this regard. In the same sense of Maturity Models, a DMMs specifically reflects the status of a company’s Digital Maturity \citep{teichert2019digital} towards DT. 
In the context of DMMs, the word ”Model” refers to the representation of an organization or business, normally obtained through a set of dimensions, while ”Maturity”  refers to a complete, perfect, or ready state of being and is the result of the progress in the development of a system \citep{teichert2019digital}.

DMMs can be descriptive, prescriptive or comparative. Descriptive DMMs are assessment models, in which the current capabilities of the entity under investigation are assessed against a given criterion \citep{becker2009developing}. In this case, the model is mainly used as a diagnostic tool. It provides a good overview of the historical development of a company, but offers little guidance on how to proceed (for planning, it is used in combination with a road-maps). Prescriptive DMMs, in contrast, are those that provide actionable plans. Prescriptive DMMs aims to identify desirable levels of maturity and guide an organization on ways to improve its measures \citep{becker2009developing}. Furthermore, comparative DMMs allow for internal or external benchmarking and offer the opportunity to compare the maturity levels of similar business units or organizations \citep{kretzschmar2021roadmap}.

A DMMs typically consists of: 
\begin{itemize}
    \item \textbf{Dimensions}: They represent key capacities of the company, e.g., culture, people, customer service, technology and data usage. As an example, the dimension “Culture \& People” can be seen as a perspective of human beings in relation to changes in the digital era, both in the role of leader-author of transformations and as an instrument in the new configurations of societies and businesses, with a focus on innovation and entrepreneurship practices transformative within organizations \citep{peixoto2019TDautomotive};
    \item \textbf{Dimension components}: Various authors emphasize that a dimension is not monolithic. Instead, dimensions are often composed of sub-dimensions \citep{barry2022benchmarking}, attributes \citep{teichert2019digital}, or components \citep{barry2023strengths} which describe the dimension in more detail. In the “Culture \& People” dimension, common components include leadership support, digital skills and openness to new technologies and change \citep{wagire2021development, bibby2018defining, paavel2017plm, o2011digital}. We prefer the term “components” because, given that dimensions resemble vectors in a mathematical sense, it is more coherent to view components as the projections of the vector;
    \item \textbf{Assessment questions}: Questions operationalize the evaluation of the maturity in a given dimension. These questions are often presented as statements to which the participants respond using LIKERT scales. For instance, one of the questions used to assess Culture \& People maturity, as outlined by \citet{peixoto2019TDautomotive}, is: “Our Innovation and DT initiatives are carried out by multidisciplinary teams from different sectors of the organization.”
    \item \textbf{Levels}: They are assigned to each dimension according to the evaluation of the participants. Levels can be quantitative (1 to 5) qualitative such as “Not Started,” “Starting,” “Enabling,” “Integrating,” “Optimizing,” and “Pioneering” as proposed by \citet{valdez2016digital}. An overall digital maturity level for the company can be obtained by performing a weighted sum of the levels of each dimension.
\end{itemize}

\begin{figure}[ht]
\centering
\begin{minipage}{0.90\textwidth}
\includegraphics [width=\textwidth]{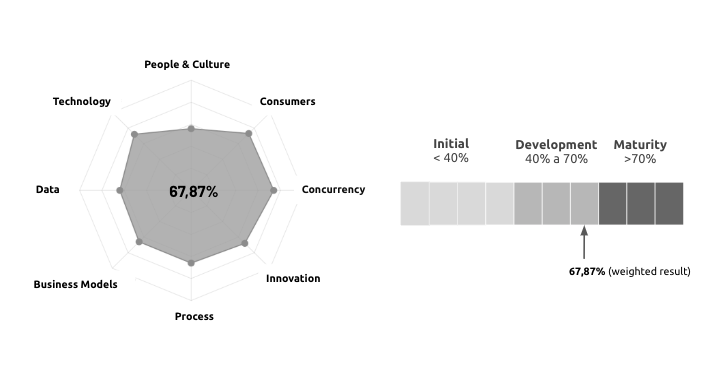}
\caption{An Example of DMMs Dimensions \& Levels Measurement Results
}\label{fig1}
\end{minipage}
\end{figure}

To exemplify, Figure \ref{fig1} presents a fictitious DMMs with eight dimensions, the levels of each dimension and also the overall level of digital maturity.

\section {Research problem}
Despite their widespread use by practitioners as tools for guiding DT, DMMs remain largely unstandardized and underdeveloped, highlighting the need for greater scholarly attention. Early in our studies, later confirmed by our systematic mapping, we have identified a few issues. 

First, there is a high degree of divergence in what is considered a relevant dimension among current DMMs. \citet{teichert2019digital} conducted an extensive systematic review of literature of 22 DMMs and identified 125 distinct dimensions across these models. However, only 41 dimensions were common to all, making comparative analysis across different models exceedingly difficult. The remaining dimensions exhibited considerable variation, not only in content but also in terminology, further exacerbating the challenge of comparing and synthesizing results from different models \citep{teichert2019digital}. This lack of agreement hampers the ability of organizations to assess digital maturity consistently and also limits the academic community’s ability to conduct rigorous comparative studies.

The second issue relates to the poor definition and vague articulation of dimensions in many DMMs. As also noted by \citet{teichert2019digital}, many models lack consistency and clarity in what each dimension aims to measure, leading to ambiguity in interpretation and application: "With regards to the design of the models, an inconsistency of levels and characteristics describing digital maturity can be found across all examined models". Other authors also corroborate this problem, such as \citet{silva2024digital}.
Without clear definitions, the models risk being interpreted differently by various organizations and researchers, undermining the reliability of their use and the validity of findings derived from them. This inconsistency is compounded by the sheer diversity of models and dimensions available, contributing further to the confusion.

The third issue, closely tied to the previous one, is that the components of a dimension are not always explicit. As \citet{barry2022benchmarking} note, “in complex domains, the identification of sub-components (sub-dimensions) is recommended. This facilitates the development of assessment questions and enriches the maturity results”. Unclear identification of components not only complicates the formulation of questions but may also result in different questions for the same dimension.

These issues—lack of consensus on important dimensions and the inconsistent definition of those dimensions and their components—are closely intertwined. The ambiguity in how dimensions are defined and applied likely leads to the significant variation in the dimensions themselves across models, as practitioners and scholars may interpret vague or ill-defined constructs in different ways. As a result, efforts to conduct comparative analyses of DMMs, or to create a unified framework for assessing DT, become extremely difficult \citep{barry2023strengths}. The disparity among models creates difficulties in determining which DMMs best suits an organization’s specific needs, and represents a critical challenge for organizations attempting to benchmark their digital maturity against industry standards or competitors.

\section {State of the Art on DMMs}
Prior to our study, we identified eight systematic analyses focusing on Digital Maturity Models (DMMs). These analyses are secondary studies that examine and synthesize primary studies proposing one or more DMMs. By systematic, we refer to a structured and methodical approach to the identification, selection, and analysis of research evidence. Such a methodology is essential to ensure rigor, transparency, and completeness in the processes of searching, screening, and categorizing the relevant studies \citep{christou2024systematic, petersen2015guidelines}. 

Notably, two of these eight studies \citep{rakoma2021review, maganjo2022measurement} are master’s dissertations that focus specifically on identifying DMMs for a narrow sector: shipping companies. Concerning the remaining six studies, they each focus on different aspects of DMMs, and none share our objective, which is to clarify the core dimensions of DMMs or address our second research question on how each dimension is described in the studies. Moreover, three of these six studies are limited to DMMs devoted to small and medium-sized enterprises (SME).

The study by \citet{ochoa2021digital} analyzes 18 DMMs in light of enterprise architecture. Their study aims to identify the architecture layers included in each dimension of these DMMs. They employed the ArchiMate 3.0 framework, a language devoted to describe, analyze, and communicate different aspects of the enterprise architecture, to determine the architecture layers. Therefore, its scope is quite different from ours.

The study by \citet{thordsen2020measure} critically evaluates 17 DMMs based on their adherence to established academic standards for measurement. Their focus is on the validity of these models in measuring a company's degree of digitalization, examining aspects like generalizability and theory-based interpretation. They conclude that “most of the identified models do not conform to the established evaluation criteria”. While this study assesses the quality of measurement in DMMs, it does not engage with the core dimensions or how they are described, which is the focus of our research.

The study by \citet{williams2022applicable} investigates 12 DMMs specifically designed for SMEs. They seek to answer three research questions: What is the current landscape of DMMs for SME? What capabilities are relevant to existing DMMs for SME? What is the focus of the capabilities of SME with high digital maturity? Their main finding is that despite the significance of DMMs across several industries, it remains limited by the lack of procedural properties and suitable SME-specific capabilities. Then, their research is centered on industry-specific needs and procedural limitations, rather than addressing the broader core dimensions of DMMs.

In a similar vein, \citet{viloria2022review} conducted a systematic review to select and analyze 5 DMMs used in SME. They conclude that the majority of the DMMs require experts accompanying the diagnostic process, preventing the company from self-applying DMMs. Although this study addresses the practical applicability of DMMs in SME, it does not aim to clarify or analyze the core dimensions of DMMs, which is our study's primary focus. By the way, the fact that an expert is necessary to guide the process somehow reinforces the necessity of a better definition of DMMs dimensions and components.

The study by \citet{cognet2023systematic} analyzed 13 DMMs that were published before 2020. These models were selected based on their availability, citation level, and validation through various test cases. The study aims to create a systematic framework that highlights the scope, similarities, and differences among various DMMs. They conclude that while no single DMMs comprehensively covers all dimensions, combining elements from multiple models provides a more complete assessment. The study suggests that future work should focus on developing a unified maturity model that synthesizes all dimensions from the evaluated models, which would enable organizations to tailor their assessments based on their specific needs.

Lastly, the study by \citet{williams2019digital} examines 25 DMMs focused on SMEs. They seek to answer two research questions: What does an empirical and conceptual DMMs for SME look like? How can DMMs be validated? The main finding is the discovery of “six essential maturity model dimensions for SME”: strategy, products/services, technology, people and culture, management, and processes. Like the other SME-focused studies, this paper is concerned with practical applications for small businesses, rather than engaging in a deep analysis of the dimensions of DMMs. Thus, it too differs from our work, which seeks to clarify the underlying DMMs dimensions across various sectors.

Finally, besides the differences in the scope and purpose of these studies, we would like to draw attention to the limited number of DMMs analyzed in each of them : 18 by \citet{ochoa2021digital}, 17 by \citet{thordsen2020measure}, 5 by \citet{viloria2022review}, 12 by \citet{williams2022applicable}, 13 by \citet{cognet2023systematic} and 25 \citet{williams2019digital}, respectively. Besides, three of the studies are focused only on DMMs to be used by SME, while we seek for a broader perspective of DMMs across different sizes of companies.

\section{Method: systematic mapping}
The specific steps undertaken in our search for DMMs literature and model analysis are elaborated upon in the subsequent sections. We followed the method proposed by \citet{keele2007guidelines}.

In the context of this research, our aim is to scrutinize extensively the existing body of literature to extend the existing knowledge over DMMs.

\subsection{Search and Selection Process}
\begin{figure}[ht]
\centering
\begin{minipage}{0.70\textwidth}
\includegraphics [width=\textwidth]{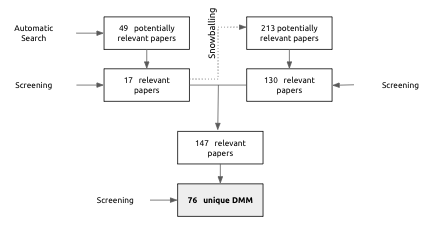}
\caption{Search and Selection Process}
\label{fig2}
\end{minipage}
\end{figure}

To collect relevant literature, we carried out an iterative search and selection process in five steps, as shown in Figure \ref{fig2}. After a brief analysis of technical terms related to DMMs and simulating different combinations of search strings, we settled on the search string “Digital Maturity Models”. In April 2023, we used this search string to perform an automatic search on title words (using the “allintitle:” operator) on Google Scholar. This initial step yielded 49 papers. 

After excluding duplicates and inaccessible papers, we conducted a first screening using the following exclusion criteria: papers not written in English, DMMs not devoted to companies and papers that do not answer any of our research questions. This left us with 17 papers for further investigation through snowballing.

Through backward snowballing of these papers, we identified additional references on related DMMs, aggregating them into a candidate list of 213 papers. During the screening phase, applying the same criteria of the first screening, we eliminated 70 duplicates and 13 that were not accessible, resulting in 130 papers to the eligibility phase. 

Finally, we performed a last more detailed screening on the remaining 147 papers to extract 76 unique DMMs. In other words, when a DMMs appeared in more than one paper, we kept only one version of it, the one described with more details.

For performed screenings, two researchers worked independently analyzing the papers, and applying the same exclusion criteria. Meetings were held to resolve any disagreements, with assistance from two more experienced researchers.

\subsection{Data Extraction and Analysis}
A spreadsheet was used to record the information from the selected papers. Two researchers further piloted the spread sheet, checking for consistency. The retrieved data includes:
\begin{itemize}
    \item \textbf{Reference}: An identification of the paper it came from, model name, authors, year, publication paper type, country and written language;
    \item \textbf{Model characteristics}: Number of dimensions, dimensions name, number of maturity levels, target application sector, target company size, developers origin (academic or practitioners);
    \item \textbf{Dimensions characteristics}: textual descriptions of the dimensions; number of components and the components under each dimension;
    \item \textbf{Assessment questions}: (when available).
\end{itemize}

In a similar manner to the search and selection strategy, the retrieved data was carefully analyzed by two researchers. When uncertainties arose, two more experienced researchers were consulted to resolve conflicts.

\subsection{Data Coding}
With the data extracted, we performed the coding process that transforms raw data into preliminary constructs, representing the dimensions of DMMs. We proceeded this in two steps. In the first step, for each study, we copied the dimensions associated with the proposed DMMs. Then, following a bottom-up conceptual clustering process, we iteratively grouped these dimensions into clusters according to their naming matching. Two identical or partially identical names were grouped together. For instance, “strategy”, “sales strategy” and “IT strategy” were all grouped together under “strategy”. 

Throughout this process, careful attention was paid to maintaining consistency and coherence of the extracted data, ensuring that similar dimensions were grouped together despite variations in terminology used by different authors. In a few cases, we needed to “split” the originally proposed dimension, since it could belong at the same time to two emerging dimensions. For instance, the dimension Culture \& People as presented in \citet{wagire2021development} was split into culture dimension and people dimension. 

Finally, we assigned a label for each cluster, representing thus a preliminary construct corresponding to a DMMs dimension.

In a similar manner, for each of the dimensions that were identified, the dimension components were extracted in order to provide a more detailed description of the dimension. 

Initially, the components mentioned for each dimension identified in the study were copied. Then, following a conceptual grouping process, the components were iteratively grouped into clusters according to their matching names and concepts.

To illustrate this process, the Empowerment Through Digital Skills dimension component of Culture was grouped with other components such as Development of Digital Literacy, Digital Skills Training, Digital Skills and Qualifications, and Digital and Leadership Skills.

\subsection{Validity Threats}
In this study, to be extensive, we included DMMs from gray literature, since we would like to  ensure a comprehensive overview of the research field, including  the diverse perspectives from practitioners and industry experts, beyond data found in peer-reviewed publications. 

Due to the novelty of the subject and the lack of confidence about the best scientific bases to search such a transversal subject, we decided on Google Scholar engine, because of its ability to index several databases, and it is recommended as the most comprehensive meta-search engine \citep{martin2018google, halevi2017suitability}. We cannot guarantee that this is the best set of papers to look at, but it returned a sensible number of unique papers, so that we could snowball and proceed to the eligibility phase. The number of included papers are well beyond the number of models we found in any other literature analysis.

By snowballing, we might have introduced some bias due to the nature of the initial paper list. Indeed, we did not care if the list were well balanced by sector, size or any other company key characteristic. Nevertheless, it complies to the recommendations given by \citet{wohlin2014guidelines} for a snowball good start set: 1- papers are relevante and came from different communities; 2- the number of papers in the start set is not too small; 3- the start set covers several different publishers, years and authors; 4- the start set is formulated from keywords in the research question, although it does not take synonyms into account.

Finally, extraction and coding methods may introduce biases during the data collection process for clustering the dimensions. That is why we involved two researchers independently extracting and coding information from the selected papers, under the supervision of two senior researchers who also intervened in solving conflicts and analyzing borderline cases.

\section{Results: Demographics of the DMM} \label{sec6}
\begin{figure}[ht]
\centering
\begin{minipage}{0.60\textwidth}
\includegraphics [width=\textwidth]{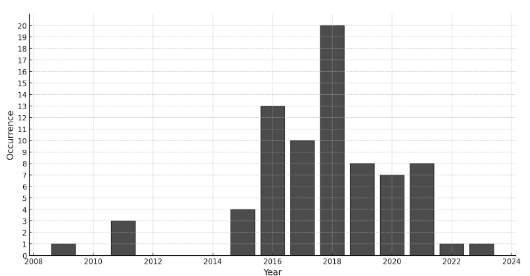}
\caption{DMMs per Year}
\label{fig3}
\end{minipage}
\end{figure}

In this section we cover general data regarding the 76 analyzed DMMs. Figure \ref{fig3} shows the quantity of analyzed papers per year of publication, showing that most of the studies were published from 2015 to 2021.

\begin{figure}[ht]
\centering
\begin{minipage}{0.65\textwidth}
\includegraphics [width=\textwidth]{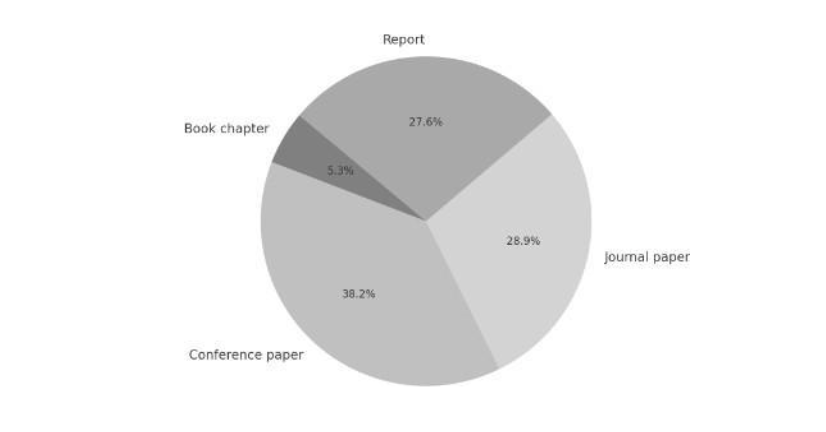}
\caption{Occurrence of Document Types}
\label{fig4}
\end{minipage}
\end{figure}

We classified the selected articles into journal articles, conference articles, book chapters and reports (those published by industry or consulting firms). Most of them, approximately 67\% of the relevant publications are found in academic conferences or journals, while only 28\% originate from reports. See Figure \ref{fig4}.

\begin{figure}[h!]
\centering
\begin{minipage}{0.70\textwidth}
\includegraphics [width=\textwidth]{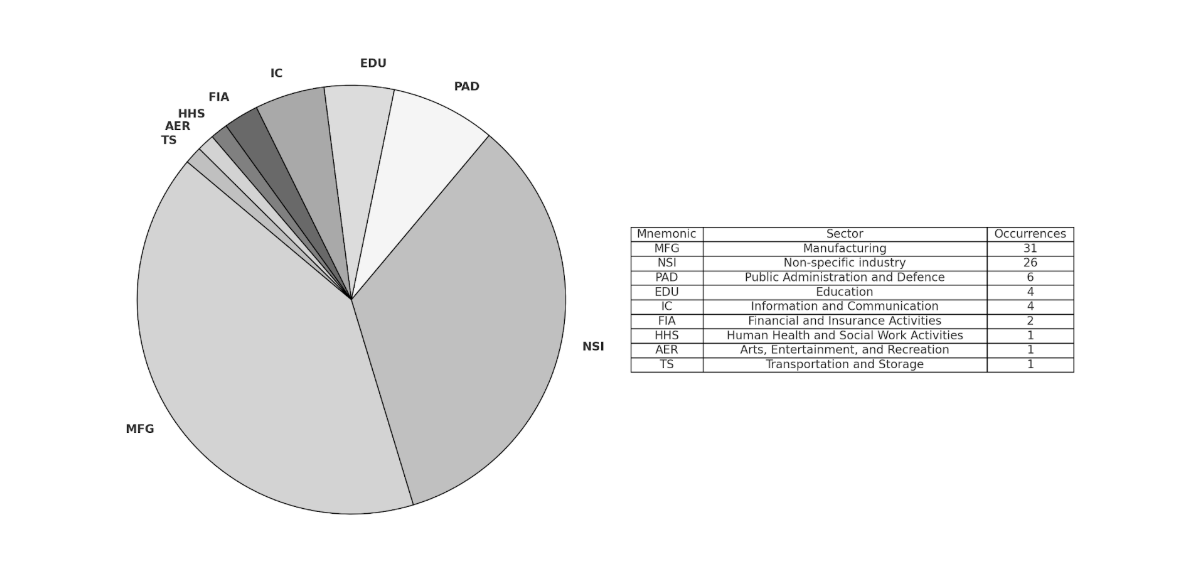}
\caption{Economic sectors \citep{nace_rev2}  to which the DMMs are devoted}
\label{fig5}
\end{minipage}
\end{figure}

Figure \ref{fig5} presents the economic sectors where DMMs are intended to be applied, with the sectors classified according to the \citet{nace_rev2}.  Note that a significant portion of the DMMs (31 out of 76) is devoted to manufacturing.

Yet concerning DMMs demographics, 12 (15,8\%) of them were devoted to Small and Medium-sized Enterprise (SME), while the remaining were conceived for a broader audience.

\section{Results: responses for RQ1}
\begin{figure}[h!]
\centering
\begin{minipage}{0.60\textwidth}
\includegraphics [width=\textwidth]{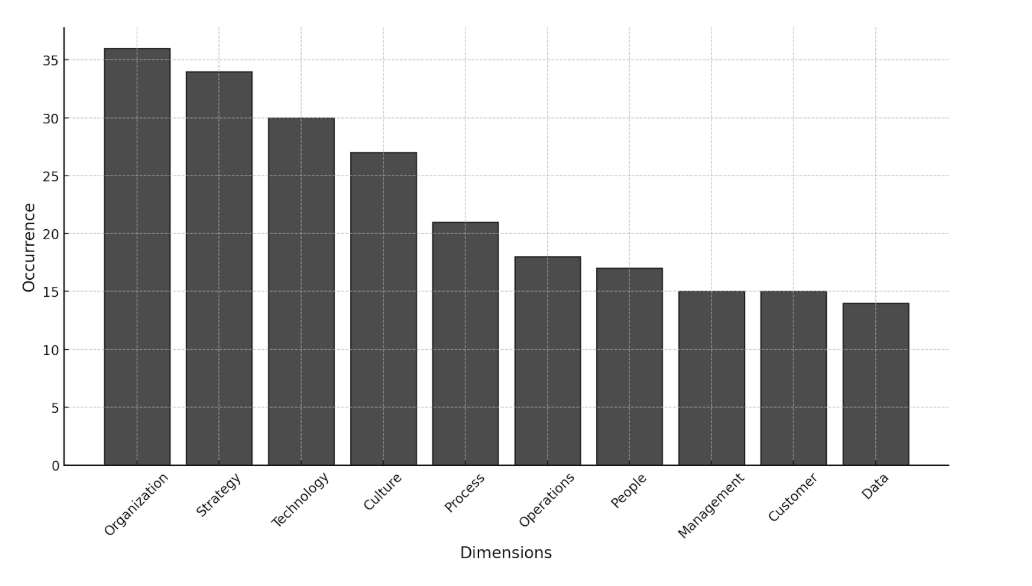}
\caption{Dimensions Occurrence}
\label{fig6}
\end{minipage}
\end{figure}

This section presents the findings from our comprehensive literature mapping. We found that the most frequent dimensions by word occurrence in the dimension names are: Organization (36), followed by Strategy (34), Technology (28), Culture (27) and Process (21). See Figure \ref{fig6}.

We also observed that models with fewer dimensions may fuse more than one dimension in just one, e.g., Strategy \& Organization \citep{agca2017industry}; Culture, People \& Organization \citep{TMForumDMM}. Then, it is imprecise to assume that dimensions with different names in various models are distinct, and that dimensions with the same name across different models are measuring the same construct.

\begin{figure}[h!]
\centering
\begin{minipage}{0.60\textwidth}
\includegraphics [width=\textwidth]{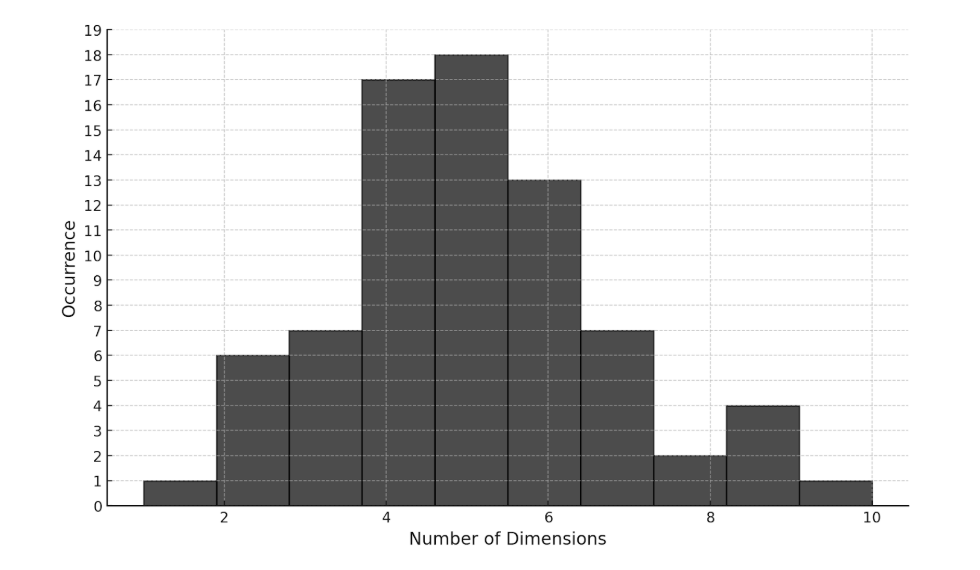}
\caption{Number of Dimensions per DMM}
\label{fig7}
\end{minipage}
\end{figure}

As illustrated in Figure \ref{fig7}, we also found that 63\% of the models consist of 4 to 6 dimensions. The model with the fewest dimensions has just 01 dimension \citep{guarino2020digital}, while the model with the most number of dimensions has 10 dimensions \citep{salviotti2019strategic}. They are both from Italy. The average number of dimensions and the mode across the models reviewed is 5. This indicates that most DMMs prefer to represent companies using a moderate range of dimensions.

\section{Results: responses for RQ2}
Understanding what is being measured is essential for effectively managing the DT process. Our research highlights the lack of consistency and clarity in the DMMs literature regarding dimension definitions. This weakness in maturity models is also noted by other authors.

\textit{”... 22 included DMMs comprise 125 maturity dimensions, out of which 41 are similar across all models and 84 have quite different and unique denominations, which makes comparability impossible.”} \citet{teichert2019digital}

In the following subsections, we provide further clarification on the 10 most frequent dimensions: Organization, Strategy, Technology, Culture, Process, Operations, People, Management, Customer and Data. We describe how they are named, highlight the components, and propose a definition for each.

\subsection{The Organization Dimension}
The Organization dimension in the context of DT can be described as follows:

\subsubsection{Construct naming}
The Organization dimension is named variously across different DMMs as follows [dimension (occurrence)]: Organization (12), Strategies \& organizations (4), Organization \& culture (3), Organizational structure (3), Culture, people \& organization  (2), Business \& Organization strategy  (1), IT management/organization (1), Company culture \&  organization  (1), Organization  \&  talent  (1), Organization \& process (1), Organizational alignment (1), Organizational capability (1), Organizational resilience (1), Organizational maturity (1), Organizational strategy (1), Business and Organization Strategy (1), Organization, employees and digital culture (1);

\subsubsection{Dimension components} 
See Organization components in Table \ref{tab:ORGcomponents}

\begin{table}[ht]
\centering
\caption{Components of Organization}
\label{tab:ORGcomponents}
\scriptsize{
\begin{tabular}{|p{3cm}|p{7cm}|p{5cm}|}
\hline
\textbf{Components} & \textbf{Description} & \textbf{DMM} \\
\hline
Strategic Alignment & Focuses on aligning organizational strategies, structures, and operations to support DT goals. It includes elements that integrate strategic planning, business model adaptation, and governance to achieve a cohesive approach to Industry 4.0 and digital maturity. & \cite{wagire2021development}, \cite{catlin2015raising}, \cite{ustundag2017industry}, \cite{AndersonWilliam2018DigitalMaturity}, \cite{agca2017industry}, \cite{vanboskirk2017digital}, \cite{zeller2018acatech} \\
\hline
Cultural and Leadership Transformation & Emphasizes the importance of leadership, culture, and mindset in enabling and sustaining DT. It focuses on creating a digital-ready culture that is adaptive, innovative, and collaborative. & \cite{blatz2018maturity}, \cite{lin2020dynamic}, \cite{almasbekkyzy2021digital}, \cite{gimpel2018structuring}, \cite{Newman2017DigitalMaturityModel}, \cite{AndersonWilliam2018DigitalMaturity}, \cite{TMForumDMM}, \cite{vanboskirk2017digital}, \cite{bandara2019model} \\
\hline
Workforce Enablement & Focuses on empowering employees with the skills, tools, and frameworks necessary to thrive in a digital organization. It emphasizes workforce readiness, training, and enablement to support transformation efforts. & \cite{ivanvcic2019mastering}, \cite{almasbekkyzy2021digital}, \cite{gill2016digital}, \cite{ustundag2017industry}, \cite{Newman2017DigitalMaturityModel}, \cite{AndersonWilliam2018DigitalMaturity}, \cite{nerima2021towards}, \cite{TMForumDMM} \\
\hline
Process and Operational Excellence & Emphasizes the optimization of processes, workflows, and operational capabilities to support DT. It includes agility, lifecycle management, and process innovation to drive efficiency and effectiveness. & \cite{chonsawat2019development}, \cite{gill2016digital}, \cite{gimpel2018structuring}, \cite{amaral2021framework}, \cite{paavel2017plm}, \cite{nerima2021towards} \\
\hline
Technological Integration and Infrastructure & Focuses on integrating digital technologies, ensuring IT readiness, and building the infrastructure needed for Industry 4.0 transformation. It covers the adoption of innovative technologies and the alignment of IT and business strategies. & \cite{isaev2018evaluation}, \cite{berghaus2016stages}, \cite{wagire2021development}, \cite{klotzer2017toward}, \cite{vanboskirk2017digital} \\
\hline
\end{tabular}
}
\end{table}

\subsubsection{Proposed construct definition}
The Organization dimension represents the structural, cultural, and operational readiness of a company to support and sustain DT. It encompasses the alignment of internal frameworks, governance, and resources with strategic goals, fostering agility, collaboration, and adaptability. This dimension addresses how leadership, workforce capabilities, and processes are structured to drive innovation, integrate technology, and maintain resilience in a rapidly changing digital environment.

\subsection{The Strategy Dimension}
The Strategy dimension in the context of DT can be described as follows:

\subsubsection{Construct naming}
The Strategy dimension is named variously across different DMMs as follows [dimension (occurrence)]: Strategy (17), Strategy \& organization  (4), Strategy \& leadership (2), Organizational strategy  (1),  Business  \&  Organization  strategy (1), Business strategy (1), Digistrategy (1), Technology strategy (1), Strategy \& policy (1), Strategy \& planning (1), Strategy development (1), Developing a digitally enabled growth strategy \& mindset (1) and Getting the strategy right (1);

\subsubsection{Dimension components}
See Strategy components in Table \ref{tab:STRcomponents}.

\begin{table}[ht]
\centering
\caption{Components of Strategy}
\label{tab:STRcomponents}
\scriptsize{
\begin{tabular}{|p{3cm}|p{7cm}|p{5cm}|}
\hline
\textbf{Components} & \textbf{Description} & \textbf{DMM} \\
\hline
Strategic Alignment & Focuses on aligning digital strategy with organizational goals, ensuring coherence across business models, operational practices, and leadership initiatives. & \cite{isaev2018evaluation}, \cite{gatziu2018fhnw}, \cite{paavel2017plm}, \cite{ustundag2017industry}, \cite{kiron2016aligning}, \cite{nerima2021towards}, \cite{o2011digital}, \cite{bandara2019model} \\
\hline
Transformation Roadmap & Centers on the creation and execution of detailed roadmaps for DT, integrating technology, resources, and strategic goals. & \cite{berghaus2016stages}, \cite{wagire2021development}, \cite{aramburu2021digital}, \cite{klotzer2017toward}, \cite{Lichtblau2017Industrie4Readiness}, \cite{scremin2018towards} \\
\hline
Innovation and Competitive Advantage & Emphasizes innovation as a driver for competitive advantage, leveraging technology, collaboration, and customer-centric initiatives. & \cite{chonsawat2020defining}, \cite{bibby2018defining}, \cite{almasbekkyzy2021digital}, \cite{Newman2017DigitalMaturityModel}, \cite{agca2017industry} \\
\hline
Governance and Leadership & Involves leadership’s role in driving strategy execution, governance practices, and the integration of key performance metrics. & \cite{ifenthaler2020development}, \cite{rossmann2018digital}, \cite{ustundag2017industry}, \cite{o2011digital} \\
\hline
Operational Integration & Focuses on integrating operational processes with digital strategies, ensuring adaptability, and leveraging technology to enhance efficiency. & \cite{wagire2021development}, \cite{Lichtblau2017Industrie4Readiness}, \cite{AndersonWilliam2018DigitalMaturity}, \cite{bandara2019model} \\
\hline
Resource Optimization & Encompasses the allocation and management of resources to support DT, ensuring sustainability and scalability. & \cite{seitz2018digital}, \cite{almasbekkyzy2021digital}, \cite{nerima2021towards}, \cite{o2011digital} \\
\hline
\end{tabular}
}
\end{table}

\subsubsection{Proposed construct definition}
The Strategy dimension in DT refers to the deliberate alignment of an organization’s vision, goals, and initiatives with digital technologies and capabilities to enable sustainable growth, innovation, and competitive advantage. This dimension encompasses the development, execution, and continuous evaluation of a strategic roadmap that integrates digital initiatives with the overall business model, organizational culture, and operational processes. The Strategy dimension serves as the foundation for DT by ensuring that technological advancements are purposefully integrated into the organization’s broader objectives. It helps organizations prioritize initiatives, allocate resources effectively, and build resilience in an increasingly digital economy.

\subsection{The Technology Dimension}
The dimension Technology in the context of DT can be described as follows:

\subsubsection{Construct naming}
The Technology dimension is referred to by various names across different DMMs as follows [dimension (occurrence)]: Technology (17), Digital technology (2), Cross-sectional technology criteria (1), Information technology (1), Educational technology (1), Technology and resources (1), Smart manufacturing technology (1), Technology driven process (1), Technology strategy (1), Technology transfer \& Service to society (1), Product and services oriented technology (1), Industry 4.0 base technology (1) and Integration of digital technology (1).

\subsubsection{Dimension components} 
See Technology components in Table \ref{tab:TECcomponents}.

\begin{table}[ht]
\centering
\caption{Components of Technology}
\label{tab:TECcomponents}
\scriptsize{
\begin{tabular}{|p{3cm}|p{7cm}|p{5cm}|}
\hline
\textbf{Components} & \textbf{Description} & \textbf{DMM} \\
\hline
Data and Analytics-Driven & Focuses on data collection, management, analysis, and the use of analytics to support decision-making and operational efficiency. & \cite{colli2018contextualizing}, \cite{chonsawat2019development}, \cite{ivanvcic2019mastering}, \cite{wagire2021development}, \cite{almasbekkyzy2021digital}, \cite{chonsawat2020defining}, \cite{leyh2016simmi} \\
\hline
Integration and Interoperability & Integrates systems, technologies, and processes to improve connectivity and seamless operations. & \cite{ifenthaler2020development}, \cite{chonsawat2019development}, \cite{sjodin2018smart}, \cite{paavel2017plm}, \cite{scremin2018towards}, \cite{Newman2017DigitalMaturityModel} \\
\hline
Advanced and Emerging Technologies & Emphasizes the adoption and utilization of advanced and emerging technologies, including AI, robotics, and blockchain. & \cite{chonsawat2019development}, \cite{ivanvcic2019mastering}, \cite{wagire2021development}, \cite{bandara2019model} \\
\hline
Digital Process Automation & Focuses on automating business and operational processes for efficiency and consistency. & \cite{chonsawat2019development}, \cite{berghaus2016stages}, \cite{ivanvcic2019mastering}, \cite{rossmann2018digital}, \cite{paavel2017plm} \\
\hline
Security and Governance & Address IT security, governance, and risk management to ensure safe and structured DT. & \cite{wagire2021development}, \cite{leyh2016simmi}, \cite{scremin2018towards}, \cite{Newman2017DigitalMaturityModel} \\
\hline
Customer and Stakeholder Focus & Prioritizes customer interaction, satisfaction, and stakeholder collaboration using digital tools. & \cite{djurek2018assessing}, \cite{ivanvcic2019mastering}, \cite{wagire2021development}, \cite{bandara2019model} \\
\hline
Strategic and Agile Enablement & Enables flexible, strategic approaches to adopting and managing DT technologies. & \cite{berghaus2016stages}, \cite{almasbekkyzy2021digital}, \cite{AndersonWilliam2018DigitalMaturity}, \cite{vanboskirk2017digital} \\
\hline
\end{tabular}
}
\end{table}

\subsubsection{Proposed construct definition}
The Technology dimension in DT refers to the set of digital tools, systems, and processes that enable an organization to enhance operational efficiency, integrate and manage data, adopt emerging innovations, automate processes, and create value through improved connectivity, analytics, and customer engagement. It serves as the foundation for achieving strategic objectives by fostering interoperability, scalability, and security while addressing the barriers to adoption and alignment with business goals. This comprehensive definition encompasses the multifaceted role of technology within DT efforts, highlighting its pivotal role in fostering innovation, agility, and value creation.

\subsection{The Culture Dimension}
The dimension Culture in the context of DT can be described as follows:

\subsubsection{Construct naming}
The dimensions related to Culture are named variously across different DMMs, as follows [dimension  (occurrence)]:  Culture  (10),  Culture  \&  people  (4), ICT culture (3), Organization \& culture (3), Culture, people \& organization (2), Innovation culture (1), Culture \& expertise (1), A fast, agile culture (1), Company culture \& organization (1), Organization, employees \& digital culture (1).

\subsubsection{Dimension components}
See Culture components in Table \ref{tab:CULcomponents}.
\begin{table}[ht]
\centering
\caption{Components of Culture}
\label{tab:CULcomponents}
\scriptsize{
\begin{tabular}{|p{3cm}|p{7cm}|p{5cm}|}
\hline
\textbf{Components} & \textbf{Description} & \textbf{DMM} \\
\hline
Digital Access and Resources & Focuses on providing and optimizing access to ICT resources, digital tools, and technological infrastructure to support digital operations and interaction. & \cite{jugo2017development}, \cite{djurek2018assessing}, \cite{gollhardt2020development} \\
\hline
Empowerment Through Digital Skills & Emphasizes building and enhancing digital skills among employees, students, and leadership to enable effective participation in DT. & \cite{djurek2018assessing}, \cite{gollhardt2020development}, \cite{wagire2021development}, \cite{almasbekkyzy2021digital}, \cite{rossmann2018digital} \\
\hline
Collaboration and Knowledge Sharing & Promotes collaboration within and across organizational boundaries, supported by open knowledge sharing and digital teamwork. & \cite{schumacher2016maturity}, \cite{gollhardt2020development}, \cite{o2011digital} \\
\hline
Adaptability and Innovation & Encourages a culture that embraces adaptability, openness to innovation, and experimentation to drive digital growth and transformation. & \cite{blatz2018maturity}, \cite{bibby2018defining}, \cite{rossmann2018digital}, \cite{klotzer2017toward}, \cite{Deloitte2018DigitalReadiness} \\
\hline
Governance and Leadership & Focuses on establishing structured governance, role definitions, and leadership frameworks to guide digital initiatives effectively. & \cite{blatz2018maturity}, \cite{almasbekkyzy2021digital}, \cite{Newman2017DigitalMaturityModel}, \cite{AndersonWilliam2018DigitalMaturity} \\
\hline
Continuous Learning and Improvement & Cultivates a mindset of continuous improvement, iterative learning, and proactive adaptation in digital processes. & \cite{berghaus2016stages}, \cite{wagire2021development}, \cite{bibby2018defining}, \cite{rossmann2018digital}, \cite{Deloitte2018DigitalReadiness} \\
\hline
Customer-Centric Digital Strategy & Focuses on improving customer experience through digital innovation and tailoring strategies to meet customer needs effectively. & \cite{klotzer2017toward}, \cite{o2011digital}, \cite{vanboskirk2017digital} \\
\hline
Cultural Change and Engagement & Encourages a culture of transparency, openness, and proactive change to support DT. & \cite{valdez2016digital}, \cite{schuh2017industrie}, \cite{zeller2018acatech} \\
\hline
\end{tabular}
}
\end{table}

\subsubsection{Proposed construct definition}
The Culture dimension can be defined as a foundational element of DT that fosters an environment of adaptability, collaboration, innovation, and empowerment. It emphasizes openness to change, continuous improvement, shared knowledge, and a mindset that values experimentation, learning from mistakes, and proactive engagement with digital tools and practices. Culture acts as a unifying force that aligns leadership, employees, and organizational processes toward embracing digital technologies and achieving strategic goals. This definition integrates the recurring themes from the components, such as openness, collaboration, adaptability, and innovation, which are central to cultivating a supportive culture for DT.

\subsection{The Process Dimension}
The Process dimension in the context of DT can be described as follows:

\subsubsection{Construct naming}
The Process dimension is named as follows in the reviewed papers [dimension  (occurrence)]: Process  (6), Process management  (2), Value chain \& process (2), Process digitalization (1), Processing (1), Information Processing (1), Order processing (1), Organization \& process (1), Technology driven process (1), Process transformation (1), Process organization (1), Production process (1), Processes \& operations (1) and Smart business processes (1).

\subsubsection{Dimension components}
See Process components in Table \ref{tab:PROcomponents}.
\begin{table}[ht]
\centering
\caption{Components of Process}
\label{tab:PROcomponents}
\scriptsize{
\begin{tabular}{|p{3cm}|p{7cm}|p{5cm}|}
\hline
\textbf{Components} & \textbf{Description} & \textbf{DMM} \\
\hline
Process Integration and Optimization & Focuses on aligning processes across organizational units, integrating physical and digital systems, and optimizing workflows for efficiency and performance. & \cite{de2017maturity}, \cite{wagire2021development}, \cite{nerima2021towards} \\
\hline
Data-Driven and Real-Time Processes & Emphasizes leveraging real-time data analytics, monitoring, and automation to enhance decision-making and operational efficiency. & \cite{chonsawat2019development}, \cite{wagire2021development}, \cite{gokalp2017development}, \cite{ganzarain2016three} \\
\hline
DT and Innovation & Centers on adopting digital technologies, automating processes, and fostering innovation to transform traditional business models. & \cite{chonsawat2019development}, \cite{berghaus2016stages}, \cite{klotzer2017toward} \\
\hline
Agile and Flexible Processes & Focuses on enhancing process flexibility, adaptability, and responsiveness to dynamic changes in technology and market demands. & \cite{sjodin2018smart}, \cite{nerima2021towards}, \cite{ganzarain2016three} \\
\hline
Organizational Alignment and Benefits Realization & Aims to align processes, technology, and people to meet organizational objectives and realize tangible benefits from DT. & \cite{lin2020dynamic}, \cite{klotzer2017toward}, \cite{o2011digital} \\
\hline
Operational Excellence in SME & Addresses the unique challenges of small and medium-sized enterprises (SMEs) in adopting digital practices, focusing on job scheduling, maintenance, and adaptability. & \cite{mittal2018towards}, \cite{gokalp2017development}, \cite{grooss2022balancing} \\
\hline
\end{tabular}
}
\end{table}

\subsubsection{Proposed construct definition}
The Process dimension encompasses the design, execution, monitoring, and optimization of workflows and activities within an organization. It focuses on integrating physical and digital systems, leveraging real-time data analytics, fostering agility, and aligning processes with strategic objectives to enhance efficiency, adaptability, and value creation. This dimension also includes automating tasks, coordinating across value chains, and adopting advanced technologies to enable seamless operations and innovation.

\subsection{The Operations Dimension}
The Operations dimension in the context of DT can be described as follows:

\subsubsection{Construct naming}
The Operations dimension is named as follows in the reviewed papers [dimension (occurrence)]: Operations (12), Manufacturing and operations (2), Digital business operations (1), Processes \& operations (1), Company operations (1), Smart operations (1).

\subsubsection{Dimension components}  
See Operations components in Table \ref{tab:OPEcomponents}.
\begin{table}[ht]
\centering
\caption{Components of Operations}
\label{tab:OPEcomponents}
\scriptsize{
\begin{tabular}{|p{3cm}|p{7cm}|p{5cm}|}
\hline
\textbf{Components} & \textbf{Description} & \textbf{DMM} \\
\hline
Process Standardization and Automation & Related to the rationalization of operations through the standardization of processes, the reduction of variability, and the use of automation to improve efficiency and reduce manual intervention. & \cite{blatz2018maturity}, \cite{gollhardt2020development}, \cite{chonsawat2019development}, \cite{almamalik2020development} \\
\hline
Integration and Collaboration & Includes components focusing on integrating internal and external systems and promoting cross-functional collaboration to ensure cohesive operations and enhanced decision-making. & \cite{schumacher2016maturity}, \cite{almamalik2020development}, \cite{gimpel2018structuring}, \cite{Lichtblau2017Industrie4Readiness} \\
\hline
Data and Analytics & Emphasizes the use of data for insights, optimization, and predictive capabilities, leveraging analytics to improve processes and outcomes. & \cite{bandara2019model}, \cite{agca2017industry}, \cite{bandara2019model} \\
\hline
Flexibility and Adaptability & Focuses on elements that enable organizations to respond quickly to changes in demand, adapt to new conditions, and maintain resilience in dynamic environments. & \cite{blatz2018maturity}, \cite{chonsawat2019development}, \cite{gimpel2018structuring} \\
\hline
Strategic and Agile Management & Covers elements aimed at aligning operations with strategic objectives, fostering agility in decision-making, and driving business transformation through innovative practices. & \cite{guarino2020digital}, \cite{AndersonWilliam2018DigitalMaturity}, \cite{bandara2019model} \\
\hline
Supply Chain and Manufacturing Optimization & Related to enhancing supply chain visibility, integrating manufacturing systems, and optimizing the production lifecycle through digital technologies. & \cite{almamalik2020development}, \cite{gimpel2018structuring}, \cite{agca2017industry} \\
\hline
\end{tabular}
}
\end{table}

\subsubsection{Proposed construct definition}
In the context of DT, the Operations dimension can be defined as the systematic integration, standardization, and optimization of business processes and activities to enhance efficiency, adaptability, and collaboration, leveraging digital technologies to drive strategic objectives and enable flexible, data-driven decision-making across supply chains, manufacturing, and service delivery. This definition emphasizes how operations play a foundational role in achieving successful DT.

\subsection{The People Dimension}
The People dimension in the context of DT can be described as follows:

\subsubsection{Construct naming}
The People dimension is named as follows in the reviewed papers [dimension (occurrence)]: People (10), People \& culture (4), Culture, people, \& organization (2) , People capability (1).

\subsubsection{Dimension components}
See People components in Table \ref{tab:PEOcomponents}.

\begin{table}[ht]
\centering
\caption{Components of People}
\label{tab:PEOcomponents}
\scriptsize{
\begin{tabular}{|p{3cm}|p{7cm}|p{5cm}|}
\hline
\textbf{Components} & \textbf{Description} & \textbf{DMM} \\ \hline
Leadership and Organizational Alignment & Focus on leadership, governance, organizational culture, and alignment of roles and responsibilities to support DT initiatives. It highlights the importance of leadership support, effective governance structures, and an adaptive culture. & \cite{wagire2021development}, \cite{mittal2018towards}, \cite{Newman2017DigitalMaturityModel}, \cite{TMForumDMM} \\ \hline
Workforce Skills and Development & Related to the development, enhancement, and application of employee skills. It emphasizes lifelong learning, technical training, skill acquisition, and balancing soft and technical skills to prepare the workforce for DT. & \cite{schumacher2016maturity}, \cite{chonsawat2019development}, \cite{sjodin2018smart}, \cite{amaral2021framework}, \cite{Deloitte2018DigitalReadiness}, \cite{kiron2016aligning} \\ \hline
Knowledge Sharing and Collaboration & Focuses on fostering a culture of collaboration and knowledge sharing within and across teams. It includes practices like mentorship, networking, collaboration, and creating environments that encourage employees to share expertise and learn collectively. & \cite{ivanvcic2019mastering}, \cite{carvalho2019maturity}, \cite{o2011digital} \\ \hline
Cultural Adaptation and Employee Engagement & Emphasizes the importance of fostering an inclusive, adaptable culture that engages employees emotionally and intellectually. It includes aspects like motivation, passion, innovation openness, and the ability to adapt to continuous improvement practices. & \cite{wagire2021development}, \cite{bibby2018defining}, \cite{carvalho2019maturity}, \cite{Deloitte2018DigitalReadiness}, \cite{kiron2016aligning} \\ \hline
Customer-Centric Focus & Involves aligning employee skills and organizational practices to meet customer needs and improve customer experiences. It includes customer feedback, customer journey optimization, and creating collaborative environments for customer engagement. & \cite{mittal2018towards}, \cite{o2011digital} \\ \hline
Digital and Technical Integration & This cluster deals with integrating technical expertise, digital tools, and innovative solutions into organizational practices to achieve higher levels of maturity. It includes the development of digital literacy, use of digital assets, and preparation for a digital workplace. & \cite{rossmann2018digital}, \cite{carvalho2019maturity}, \cite{Newman2017DigitalMaturityModel} \\ \hline
\end{tabular}
}
\end{table}

\subsubsection{Proposed construct definition}
The People dimension in the context of DT encompasses the capabilities, skills, attitudes, and collaborative behaviors of employees and leaders that drive DT within an organization. It focuses on equipping individuals with the necessary technical and soft skills, fostering a culture of continuous learning, collaboration, and adaptability, and aligning workforce roles and responsibilities with digital strategies. This dimension emphasizes leadership support, workforce enablement, and engagement while addressing challenges like resistance to change, skill gaps, and cultural alignment to ensure successful adoption of digital technologies and innovative practices. This dimension is vital for enabling an organization's transition into a digitally mature entity by integrating human and technological elements effectively.

\subsection{The Management  Dimension}
The Management dimension in the context of DT can be described as follows:

\subsubsection{Construct naming}
The Management dimension is named as follows in the reviewed papers [dimension (occurrence)]: Planning, management \& leadership (2), IT Management/Organization (1), ICT Project Management (1), ICT Process Management (1), Transformation management intensity (1), Transformation management (1), Management \& control (1), Asset management (1), Application management (1), Performance management maturity (1) Human resource management (1).

\subsubsection{Dimension components}
See Management components in Table \ref{tab:MANcomponents}.

\begin{table}[ht]
\centering
\caption{Components of Management}
\label{tab:MANcomponents}
\scriptsize{
\begin{tabular}{|p{3cm}|p{7cm}|p{5cm}|}
\hline
\textbf{Components} & \textbf{Description} & \textbf{DMM} \\ \hline
Strategic Vision and Alignment & Focuses on developing a cohesive strategy, aligning organizational goals with DT efforts, and ensuring a shared understanding of the vision among stakeholders. & \cite{jugo2017development}, \cite{djurek2018assessing}, \cite{westerman2011digital}, \cite{begicevic2021assessing} \\ \hline
Governance and Leadership & Focuses on establishing robust governance structures and leadership support to oversee and manage DT initiatives. & \cite{jugo2017development}, \cite{berghaus2016stages}, \cite{gimpel2018structuring} \\ \hline
Digital Competencies and Human Resource Development & Emphasizes the importance of enhancing digital skills, fostering a digital culture, and ensuring employee readiness for DT. & \cite{djurek2018assessing}, \cite{gatziu2018fhnw}, \cite{salviotti2019strategic}, \cite{gimpel2018structuring} \\ \hline
Resource Management and Operational Efficiency & Addresses the allocation of financial, technological, and human resources to optimize DT processes. & \cite{jugo2017development}, \cite{begicevic2021assessing}, \cite{paavel2017plm}, \cite{gokalp2017development} \\ \hline
Monitoring and Performance Measurement & Focuses on evaluating the effectiveness of digital initiatives and ensuring continuous improvement through performance tracking. & \cite{jugo2017development}, \cite{berghaus2016stages}, \cite{begicevic2021assessing}, \cite{jung2016overview} \\ \hline
Digital Process and Technology Integration & Emphasizes integrating technology into business processes to ensure value realization and operational success. & \cite{gatziu2018fhnw}, \cite{begicevic2021assessing}, \cite{gokalp2017development}, \cite{o2011digital} \\ \hline
\end{tabular}
}
\end{table}

\subsubsection{Proposed construct definition}
The Management dimension in the context of DT can be defined as the strategic planning, governance, resource alignment, and leadership activities necessary to oversee and guide the integration of digital technologies. It involves establishing clear roles and responsibilities, ensuring alignment with organizational goals, managing digital competencies, monitoring performance, and optimizing the use of human, financial, and technological resources. This definition synthesizes elements such as strategic oversight, resource allocation, competency development, and performance monitoring mentioned in the components.

\subsection{The Customer Dimension}
The dimension Customer in the context of DT can be described as follows:

\subsubsection{Construct naming}
The Customer dimension is named as follows in the reviewed papers [dimension (occurrence)]: Customer (10), Customer experience (1), Digital customer access (1), Digital business models and customer access (1), Offering to the customer (1), Market \& customer access (1);

\subsubsection{Dimension components}
See Customers components in Table \ref{tab:CUScomponents}.


\begin{table}[h!]
\centering
\caption{Components of Customer}
\label{tab:CUScomponents}
\scriptsize{
\begin{tabular}{|p{3cm}|p{7cm}|p{5cm}|}
\hline
\textbf{Components} & \textbf{Description} & \textbf{DMM} \\ \hline
Insights and Analytics & Focuses on leveraging data and analytics to understand customer behaviors, preferences, and needs. It emphasizes using insights to drive decision-making and improve customer interactions. & \cite{berghaus2016stages}, \cite{gimpel2018structuring}, \cite{PwC2016Industry4}, \cite{Newman2017DigitalMaturityModel}, \cite{AndersonWilliam2018DigitalMaturity}, \cite{bandara2019model} \\ \hline
Personalized Customer Experiences & Emphasizes creating tailored experiences for customers across various channels. It includes personalization of services, content, and interactions to meet individual customer needs. & \cite{berghaus2016stages}, \cite{gimpel2018structuring}, \cite{PwC2016Industry4}, \cite{Newman2017DigitalMaturityModel}, \cite{rogers2016digital}, \cite{AndersonWilliam2018DigitalMaturity}, \cite{bandara2019model} \\ \hline
Seamless Customer Journeys & Emphasizes the importance of providing consistent, integrated, and efficient customer journeys across digital and physical touchpoints. It involves managing interactions cohesively throughout the customer lifecycle. & \cite{ivanvcic2019mastering}, \cite{gimpel2018structuring}, \cite{PwCIndustry4SelfAssessment2015}, \cite{TMForumDMM}, \cite{bandara2019model} \\ \hline
Trust and Relationship Building & Focuses on fostering long-term customer relationships through trust, transparency, and consistent interactions. It also emphasizes empowering customers to interact confidently with the organization. & \cite{berghaus2016stages}, \cite{Newman2017DigitalMaturityModel}, \cite{rogers2016digital}, \cite{AndersonWilliam2018DigitalMaturity}, \cite{TMForumDMM} \\ \hline
Collaborative Value Creation & Emphasizes collaboration between organizations and customers to co-create value. It includes involving customers in the development of products, services, and strategies. & \cite{rogers2016digital}, \cite{PwCIndustry4SelfAssessment2015}, \cite{bandara2019model} \\ \hline
\end{tabular}
}
\end{table}

\subsubsection{Proposed construct definition}
The Customer dimension in DT represents the strategic focus on understanding, engaging, and delivering value to customers through personalized experiences, data-driven insights, and seamless interactions across all touch points. It emphasizes leveraging technology and analytics to gain deep insights into customer behavior and preferences, fostering trust and loyalty, and creating collaborative ecosystems where customers actively participate in co-creating value. This dimension integrates both digital and physical channels, ensuring a cohesive customer journey and empowering customers as key stakeholders in an organization's success.

\subsection{The Data Dimension}
The Data dimension in the context of DT can be described as follows:

\subsubsection{Construct naming}
The Data dimension is named as follows in the reviewed papers [dimension (occurrence)]: Data (5), Data maturity (1), Digital data (1), Data analysis (1), Data transformation (1), Data \& analytics as core capability (1), Data governance (1), Data-driven services (1), Open data (1), Data storage and compute (1);

\subsubsection{Dimension components}
See Data components in Table \ref{tab:DATcomponents}.

\begin{table}[h!]
\centering
\caption{Components of Data}
\label{tab:DATcomponents}
\scriptsize{
\begin{tabular}{|p{3cm}|p{7cm}|p{5cm}|}
\hline
\textbf{Component of Data} & \textbf{Description} & \textbf{DMM} \\ \hline
Data Governance and Security & Focuses on the establishment of policies, frameworks, and processes to ensure the secure, ethical, and effective management of data. It includes data ownership, privacy, compliance, and data security measures. & \cite{solar2017correlation}, \cite{gimpel2018structuring}, \cite{gokalp2017development}, \cite{TMForumDMM} \\ \hline
Data Integration and Processing & Emphasizes the integration of data from diverse sources, ensuring data consistency, quality, and preparation for advanced analytics and decision-making. & \cite{remane2017digital}, \cite{gimpel2018structuring}, \cite{chonsawat2020defining}, \cite{PwC2016Industry4}, \cite{nerima2021towards} \\ \hline
Advanced Data Analytics and Insight & Focuses on leveraging data for strategic insights, predictive modeling, and automated decision-making to create business value. & \cite{carvalho2019maturity}, \cite{PwC2016Industry4}, \cite{gokalp2017development}, \cite{rogers2016digital} \\ \hline
Value Realization and Data-Driven Services & Emphasizes using data for enhancing customer experiences, creating new revenue streams, and supporting innovative business models. & \cite{Lichtblau2017Industrie4Readiness}, \cite{TMForumDMM} \\ \hline
Data Storage and Computing & Addresses the infrastructure required to store, process, and compute data effectively, focusing on legacy system upgrades, real-time capabilities, and scalability. & \cite{chonsawat2020defining}, \cite{weber2017m2ddm} \\ \hline
Data Visualization and Representation & Focuses on the representation of data for clear communication and decision-making, enabling businesses to derive actionable insights. & \cite{nerima2021towards} \\ \hline
\end{tabular}
}
\end{table}

\subsubsection{Proposed construct definition}
Data is the foundational dimension that encompasses the processes, technologies, and governance mechanisms required for collecting, integrating, processing, analyzing, storing, and securing information. It serves as the backbone for enabling informed decision-making, driving innovation, enhancing operational efficiency, and creating value through data-driven services. This dimension includes key aspects such as governance, security, analytics, real-time accessibility, and the infrastructure necessary to ensure the ethical, effective, and strategic utilization of data across organizational and customer touch-points. 

\section{Other Findings about Dimensions}
We observed that a significant number of models consolidate dimensions, e.g. Organization \& Strategy and Culture \& People, potentially obscuring their distinct components. To address this, we conducted a non-exhaustive examination of similarities between dimensions, revealing several shared components. Notably, these components are often evident across multiple dimensions, particularly in areas related to organizational readiness, alignment, and integration. Consequently, despite variations in nomenclature, many dimensions exhibit a high degree of overlap, often measuring similar aspects. This overlap contributes to complexity, as it becomes unclear whether dimensions with different names assess the same construct or represent entirely distinct concepts.

Our analysis of similarities between dimensions and their components focused solely on the correspondence between nomenclature and component descriptions. This approach, while not exhaustive, opens possibilities for identifying further connections between dimensions. By doing this, we were also able to identify some distinct components within certain dimensions. Some of these similarities and distinctions are depicted in the following paragraphs. 

The alignment between organizational structure and strategy is a topic that recurs throughout this discussion. Both Strategy and Organization dimensions emphasize aligning organizational structures and leadership to support DT goals. For example, the Strategy dimension is recognized as referring to the alignment of an organization's vision, goals and initiatives with digital technologies and capabilities to enable sustainable growth, innovation and competitive advantage. Organization, in turn, emphasis the alignment of the organization focusing on internal frameworks, governance and resources with strategic goals.

The importance of leadership as a driver for DT is a key component in some dimensions such as Organization, Strategy, Management and Culture. The Strategy and Organization dimensions evaluate whether leadership is effectively aligning the company’s strategic goals with DTs. The Management dimension assesses the extent to which leadership is able to coordinate digital initiatives \citep{jugo2017development}. The Culture dimension examines how leadership promotes cultural change, enables openness to change and fosters collaboration.

In a similar manner, dimensions such as Technology, Process and Operations frequently share common objectives, which are related to operational efficiency through process optimization. Components such as Process Integration and Automation are present in more than one of these dimensions. 

Regarding operational efficiency through process optimization, the components under the Technology dimension examine the utilization of advanced digital tools to drive automation and improve operational performance. Process reinforces this vision by focusing on the streamlining of workflows through digitization (Lin et al (2020)). Conversely, the Operations dimension seeks to automate operations in order to reduce variability and increase efficiency.

The Culture and People dimensions place a significant emphasis on change management and skill development as a core enabler.
The necessity for a shift in mindset is a key aspect of DT, as evidenced by the inclusion of the Culture component (openness to change) \cite{ifenthaler2020development}, which promotes openness, collaboration, and emotional readiness for change. The People dimension addresses resistance to change through engagement and skill-building initiatives. Focuses on how employees adjust to changes driven by digital technologies. It includes aspects like innovation openness and the ability to adapt to continuous improvement practices.

With regard to the development of competence, a strong overlap was identified between the dimensions of Culture, People and Management. The Culture dimension promotes skill development within a continuous learning environment, enhancing adaptability and collaboration. The People dimension (development of digital skills) \citep{ivanvcic2019mastering} prioritizes workforce development, technical and soft skills, and knowledge sharing. The Management emphasizes building digital competencies to prepare employees for transformation.

The customer-centric transformation is emphasized in some dimensions, with a focus on enhancing value creation through insights. The Data dimension (Value realization and data-driven services) examines how companies collect and process data for customer insights, enabling value creation and predictive solutions to better meet customer needs. The emphasis here is on the utilization of data through technology for better predictions and decisions \citep{berger2015digital}. The Customer dimension (Personalised customer experiences) emphasis creating tailored experiences for customers through personalization of services, content, and interactions to meet individual customer needs. The Technology dimension, in turn, prioritizes customer interaction, satisfaction, and stakeholder collaboration using digital tools.

These similarities reinforce the idea that dimensions are interdependent, with shared goals of achieving alignment, readiness, and transformation through leadership, technology, processes, and data. They highlight the interconnected nature of organizational, technological, and human factors in achieving digital maturity. 

Among the dimensions analyzed, a few exhibit more particular components that are distinct from others. These components are highly specialized, focusing on unique aspects of DT that are not as widely addressed in other dimensions.

The Data dimension has components that provide a unique focus on data governance \citep{gokalp2017development}, analytics and infrastructure, such as data governance and security, data integration and processing, and data storage and computing. While other dimensions (such as Technology) touch on analytics or tools, the Data dimension uniquely emphasis the governance, processing and strategic use of data as a core enabler of DT.

Culture focuses on the people and culture to drive transformation \citep{PwC2016Industry4}, particularly collective behaviors and emotional and psychological aspects of how an organization deals with change, making it a more people-centric and attitudinal approach. People and Organization, however, address the issues of talent and structure (focuses on individual skills and roles).
The Technology dimension has a distinctive focus on tools, infrastructure and cybersecurity making it more about the hardware of DT \citep{amaral2021framework} as technical enablers of transformation. 

\begin{figure}[h!]
\centering
\begin{minipage}{0.60\textwidth}
    \centering
    \includegraphics[width=\textwidth]{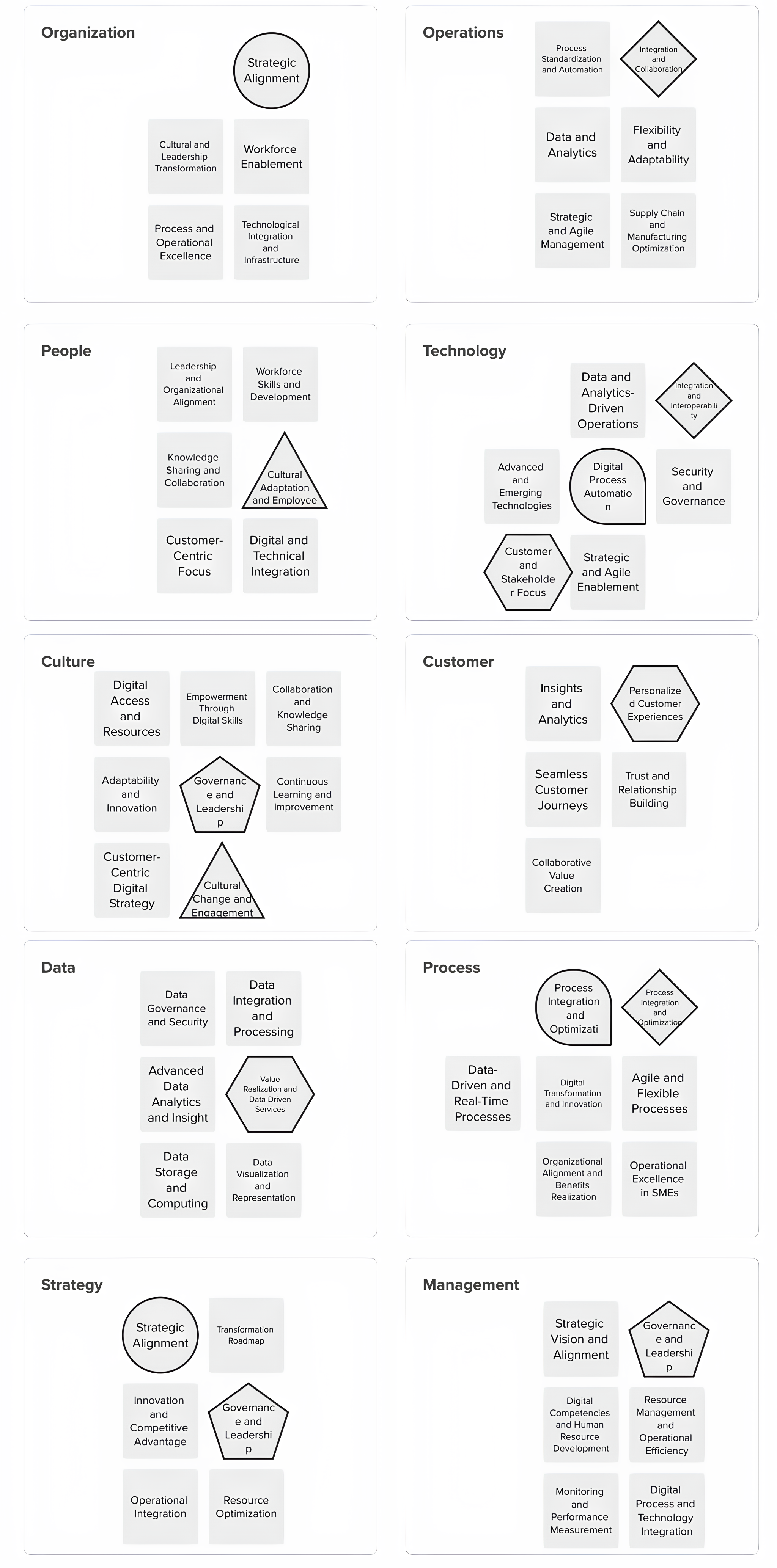}
    \caption{Dimensions components similarities and differences}
    \label{fig8}
\end{minipage}
\end{figure}

The Customer dimension is distinctive in its emphasis on enhancing customer interactions through digital means \citep{valdez2016digital}, emphasizing personalized, seamless, and efficient customer experiences \citep{berghaus2016stages}.

Some dimensions such as Data, Culture, Technology and Customer have components that contribute to a distinctive perspective on a holistic digital maturity framework, with specialized attention on different aspects of DT.

Figure \ref{fig8} illustrates the distribution of components across their respective dimensions, highlighting overlaps (represented by similar geometric patterns) while maintaining the unique contributions of each dimension.

\section{Conclusion}
This study offers an integrative framework to address key challenges in DMMs, focusing on reconciling inconsistencies and clarifying dimensions for greater academic and practical utility. By clarifying and reconciling a central issue within the DMMs — its dimensions through its components — this research enhances comprehension of the models, facilitating benchmarking and providing stronger support for organizations in navigating the complexities of DT. This comprehensive perspective equips researchers and practitioners with actionable insights to enhance the design, evaluation, and application of DMMs, ultimately advancing the field of DT and supporting organizations on their journey toward digital excellence.

By analyzing 76 DMMs through a systematic mapping process we found out that academics are significantly leading the development of DMMs (72\% of the models). The majority of studies and reports comes from Germany (n=19), the USA follows with a considerably lower number of publications (n=10). Our data do not indicate a consistent rise in interest in DMMs due to the COVID-19 pandemic. While there was a substantial number of publications during the pandemic years (more than 7 per year), this number dropped sharply afterward to approximately 2 per year.

We highlighted the fragmented nature of DMMs dimensions by analyzing their similarities and differences, and we proposed harmonized definitions for the ten most frequent dimensions: Organization, Strategy, Technology, Culture, Process, Operations, People, Management, Customer, and Data. While these definitions do not fully reconcile overlaps, they aim to provide a more consistent framework.

We also found out that most models do not precisely define each dimension, leaving practitioners to infer what is being measured. Improving the clarity of dimension definitions presents a valuable opportunity to enhance DT for practitioners and foster the development of more robust and effective DMMs by scholars. Notwithstanding the nomenclature employed, a substantial number of dimensions are designed to measure analogous constructs. A more detailed examination of the similarities between dimensions in DMMs reveals several common components. These common components are more evident across a range of dimension groups, in particular among 3 dimension groups: 1- Organization and Strategy; 2- Technology, Process and Operations; and 3- Culture, People and Management. 

Conversely, a few dimensions display more specific components that are distinct from those observed in other dimensions. The dimensions of Data, Technology and Customer contribute to distinctive perspectives on a holistic digital maturity framework, with specialized attention on different aspects of DT. These components are highly specialized, focusing on unique aspects of DT that are not as widely addressed in other dimensions.

As a direction for future research, we observed that certain dimensions, such as Organization, encompass numerous components that are also addressed across other dimensions. This observation highlights the need to re-evaluate the significance of such dimensions within the framework. Future studies could delve deeper into component similarities, potentially resulting in a more streamlined and mutually exclusive set of dimensions.

Additionally, the inclusion of Strategy as a distinct dimension deserves further consideration. If its components primarily emphasize strategic alignment and the transformation roadmap, it may indicate that the entire DMMs should be structured to assess maturity in alignment with a company's pre-selected strategy, rather than treating strategy as an independent dimension. These insights pave the way for refining DMMs frameworks and enhancing their practical utility.

\section*{Statements And Declarations}
\textbf{Competing Interests}. The authors declare that there are no actual or potential conflicts of interest related to the content of this article. They have no relevant financial or non-financial interests to disclose. Additionally, the authors confirm that they have no affiliations with or involvement in any organization or entity with a financial or non-financial interest in the subject matter or materials presented in this manuscript. The authors also state that they hold no financial or proprietary interests in any materials discussed in this article.

\textbf{Declaration of generative AI and AI-assisted technologies in the writing process}.
During the preparation of this work the author(s) used ChatGPT 4o in order to improve efficiency and clarity. After using this tool/service, the author(s) reviewed and edited the content as needed and take(s) full responsibility for the content of the publication.


\appendix
\section{Snowball source list} \label{app1}
Following is the list of articles used as sources for the snowball process: 
\cite{thordsen2020measure},
\cite{barry2022benchmarking},
\cite{barry2023strengths},
\cite{rakoma2021review},
\cite{maganjo2022measurement},
\cite{ochoa2021digital},
\cite{williams2022applicable},
\cite{viloria2022review},
\cite{williams2019digital},
\cite{grover2020comparison},
\cite{kaszas2023emergence},
\cite{bumann2019action},
\cite{cognet2019towards},
\cite{chi2022digital},
\cite{lee2022developing},
\cite{coskun2019warehouse},
\cite{cognet2023systematic}.

\section{DMMs articles list} \label{app2}
Following is the list of articles used  in this study:
\cite{rogers2016digital}, \cite{gokalp2017development}, \cite{valdez2016digital}, \cite{schumacher2016maturity}, \cite{agca2017industry}, \cite{wagire2021development}, \cite{bibby2018defining},
\cite{paavel2017plm},
\cite{o2011digital},
\cite{guarino2020digital}, \cite{salviotti2019strategic}, \cite{catlin2015raising}, \cite{ustundag2017industry}, \cite{AndersonWilliam2018DigitalMaturity}, \cite{vanboskirk2017digital}, \cite{zeller2018acatech}, \cite{blatz2018maturity},
\cite{lin2020dynamic}, \cite{almasbekkyzy2021digital}, \cite{gimpel2018structuring}, \cite{Newman2017DigitalMaturityModel}, \cite{TMForumDMM},
\cite{bandara2019model}, \cite{ivanvcic2019mastering}, \cite{gill2016digital}, \cite{nerima2021towards}, \cite{chonsawat2019development}, \cite{amaral2021framework}, \cite{isaev2018evaluation}, \cite{berghaus2016stages}, \cite{klotzer2017toward},
\cite{gatziu2018fhnw},
\cite{kiron2016aligning}, \cite{aramburu2021digital}, \cite{Lichtblau2017Industrie4Readiness}, \cite{scremin2018towards}, \cite{chonsawat2020defining}, \cite{ifenthaler2020development}, \cite{rossmann2018digital}, \cite{seitz2018digital}, \cite{colli2018contextualizing}, \cite{leyh2016simmi},
\cite{sjodin2018smart}, \cite{djurek2018assessing}, \cite{jugo2017development}, \cite{gollhardt2020development}, \cite{Deloitte2018DigitalReadiness}, \cite{schuh2017industrie}, \cite{de2017maturity}, \cite{ganzarain2016three}, \cite{mittal2018towards}, \cite{grooss2022balancing}, \cite{almamalik2020development}, \cite{carvalho2019maturity}, \cite{westerman2011digital}, \cite{begicevic2021assessing}, \cite{jung2016overview}, \cite{PwC2016Industry4}, \cite{PwCIndustry4SelfAssessment2015}, \cite{solar2017correlation}, \cite{remane2017digital},
\cite{weber2017m2ddm},
\cite{berger2015digital}, \cite{pulkkinen2018modelling}, \cite{friedrich2011},
\cite{goumeh2021digital},
\cite{PwC2017},
\cite{alalwany2007government}, \cite{chohan2020synthesizing}, \cite{kljajic2021multi}, \cite{garzoni2020fostering}, \cite{mijnhardt2016organizational}, \cite{westermann2016reference}, \cite{castor2016mesa}, \cite{ECDigitalEconomy2019}, \cite{asdecker2018development}.


\bibliographystyle{elsarticle-num-names} 
\bibliography{sn-bibliography}






\end{document}